\documentclass[twocolumn]{svjour3}

\usepackage{graphicx}
\usepackage{dcolumn}
\usepackage{bm}
\usepackage{hyperref}
\usepackage{amssymb}
\usepackage{hyperref}

\RequirePackage[normalem]{ulem} 
\RequirePackage{color}\definecolor{RED}{rgb}{1,0,0}\definecolor{BLUE}{rgb}{0,0,1} 

\begin{document}
\newcommand{\be}{\begin{equation}}
\newcommand{\ee}{\end{equation}}
\newcommand{\ba}{\begin{eqnarray}}
\newcommand{\ea}{\end{eqnarray}}
\newcommand{\Gam}{\Gamma[\varphi]}
\newcommand{\Gamm}{\Gamma[\varphi,\Theta]}
\thispagestyle{empty}

\date{\today}

\title{ Quantum correlations in  one-dimensional  Wigner molecules }

\author{ Przemys\l aw Ko\'scik, Institute of Physics,  Jan Kochanowski University\\
ul. \'Swi\c{e}tokrzyska 15, 25-406 Kielce, Poland }

\maketitle

\begin{abstract}
We studied  one-dimensional systems  formed by $N$ identical particles confined in a harmonic trap
  and
subject to an inverse power-law interaction potential $\sim|x|^{-d}$. The correlation properties of a Wigner
molecule with the lowest energy are investigated in terms of their  dependence on the number $N$ and the power $d$, including the limit as $d\rightarrow 0$. There  are  $N$-particle
Wigner molecules  with  properties such that their correlations are mainly manifested   in the $N$ lowest
 natural orbitals. The values of  the control parameters  of the system at which  such states appear are identified.
 The properties of Wigner molecules formed in the limit as $d$ approaches infinity are also revealed.

\end{abstract}

\section{Introduction}\label{1}
Soon after the birth of quantum mechanics, the study of quantum correlations began to attract significant research attention. In particular, the study of the many-body properties of various quantum composite systems is nowadays one of the most active
areas of theoretical physics \cite{1,1.2,1.3,1.4,3.1,2,2.1,2.2,2.3,3,5,6,7,7.1,12,13,szaf,ant,ant1,1.1,56,4,4.1}. Investigating quantum correlations in systems of interacting
particles trapped in external potentials is not only important in view of the context of quantum information technology
\cite{0} but also a key to improving our understanding of quantum matter.
 Various physical realizations of such systems are achieved nowadays. The advantage of artificially created systems is the possibility of controlling their properties, such
as the number of particles,  their interactions,
and the shape of the trapping potential as well.

In addition to well-known ultracold atomic gases with short-range interactions,
 systems with long-range interactions have been created in laboratories.
  In particular,  tremendous technological progress  has been made over the last two decades and opened
new perspectives for achieving the van der Waals  and dipolar interactions in polar
molecule experiments \cite{8,9}. Other well-known examples of artificially created systems exhibiting
long-range interactions are quantum dots \cite{10,10.1} and systems of  electromagnetically  trapped ions \cite{11}. Such systems exhibit a variety of strongly
correlated states. Among them is the famous Wigner phase \cite{wig}, which manifests itself in the localisation
of particles around their classical equilibrium positions.

 The  Wigner molecule-like states, formed by particles interacting via
 long-range interactions,
 have been widely studied in various
 theoretical contexts. For example, the studies
include systems with interaction potentials such as  the  Coulomb potential  \cite{12,13,szaf,ant,ant1},
the inversely quadratic
 potential \cite{1.1,56,4} and  dipole-dipole
interactions \cite{13,4.1}. The Wigner  molecules  have also been  observed  in a  variety  experiments
 \cite{11,a,b,c,d}.

The task in the studies here is to carry out a comprehensive investigation of the correlation  in
the Wigner molecule's ground state, of the one-dimensional (1D) system

 \be {\cal H}=-{1\over
2}\sum_{i=1}^N{{\partial^2\over \partial x_{i}^2}}+{\cal V} (x_{1},x_{2},...,x_{N}),\label{hamk}\ee
\be {\cal V}(x_{1},x_{2},...,x_{N})={1\over 2}\sum_{i=1}^{N} x_{i}^2
 +\sum_{i>j=1}^{N} {g\over |x_{i}-x_{j}|^d},  \label{pot1}  \ee
which  arises at positive values of $d$ in the  large interaction strength regime, $g>>1$.

In our study, we  use  a scheme based on harmonic approximation (HA), developed in  \cite{12,13},
which enables an easy determination of various   characteristics of the Wigner molecule state, such as the
density, occupation numbers, and the entanglement measures.
 The HA yields   exact results  in the limit as $g\rightarrow \infty$
 ($d>0$) \cite{13,ant,ant1,56}. However, this fails strictly within the  limit as $d\rightarrow \infty$, i.e., when   the interaction potential becomes the hard-core potential \cite{guo}.

The structure of this paper is as follows.  Section
\ref{sed}
   outlines the  formalism based on the HA
to analyse the correlation properties of the Wigner molecule states.
 In addition, it provides an effective  procedure  for the analysis of the special case as $d\rightarrow 0$.
Section \ref{results} focuses on  the question
of how the changes in the parameter $d$ affect  the   correlation in the $N$-particle Wigner molecule ground-state.  Finally, Section \ref{con} presents concluding remarks.

\section{Methods}\label{sed}
 The potential ${\cal V}$ given by Eq. (\ref{pot1}) attains its
minimum at  $N!$  points. Here we refer to the point
 $\vec{r}_{min}=(x^{c}_{1},x^{c}_{2},...,x^{c}_{N})$ with
$x^{c}_{1}<x^{c}_{2}<...<x^{c}_{N}$, $x_{i}^{c}=-x_{N-i+1}^{c}$, ($x_{{N+1\over 2}}^{c}=0$ if $N$ is odd), and
denote it by $\vec{r}^{(0)}$.
 The equilibrium positions have the form
$x_{i}^{c}=\beta_{i}^{c} g^{{1\over 2+d}}$, where $\beta_{i}^{c}$ satisfy the set of equations $\partial _{\beta_{k}}
{\cal V}^{g=1}(\beta_{1},...,\beta_{N})=0$,
\be {\cal V}^{g=1}(\beta_{1},...,\beta_{N})={1\over 2}\sum_{i=1}^{N} \beta_{i}^2
 +\sum_{i>j=1}^{N} {1\over |\beta_{i}-\beta_{j}|^d}. \label{pot} \ee
The values of  $\beta_{i}$ are known analytically only for $N = 2, 3$ \cite{ant}.
The exception is the  limit as $d\rightarrow \infty$ in which  $\beta_{i}\rightarrow$   ${(2i-N-1)/2}, i=1,...,N$. It can be  inferred that  in the limit of small $d$  in which
$$|x|^{-d}\rightarrow 1-d \mbox{ln} |x|,$$ the equilibrium positions of the particles $x_{i}^{c}$ tend to
 $x_{i}^{c}=\alpha_{i}^{c}\sqrt{d g}$, where $\alpha_{i}^{c}$
are the
solutions of the set of equations $\partial _{\alpha_{k}}{\cal V}^{d\rightarrow 0}(\alpha_{1},...,\alpha_{N})=0$,
 \be {\cal V}^{d\rightarrow 0}(\alpha_{1},...,\alpha_{N})=\sum_{i=1}^{N} \alpha_{i}^2
 -\sum_{i>j=1}^{N} \mbox{ln} (\alpha_{i}- \alpha_{j})^2. \label{pop}\ee

Within the HA the total potential (\ref{pot1}) is  approximated harmonically around $\vec{r}^{(0)}$,
\be {\cal V}(\vec{r})\approx {\cal V}(\vec{r}^{(0)})+{1\over 2!}(\vec{r}-\vec{r}^{(0)})^T \textbf{H}(\vec{r}-\vec{r}^{(0)}),\label{hrt}\ee
where $\textbf{H}$ is the so-called   Hessian matrix, which can be expressed   as \cite{13}

\be \textbf{H}=[\frac{\partial^2 {\cal V}(\beta_{1} ,...,\beta_{N} )}{\partial \beta_m\partial
\beta_k}|_{\{\beta_{i}=\beta_{i}^{c}\}}]_{N\times N}.\ee
In the particular case as $d\rightarrow 0$, we obtained

\be \textbf{H}^{d\rightarrow 0}={1\over 2}[\frac{\partial^2 {\cal V}^{d\rightarrow 0}(\alpha_{1} ,...,\alpha_{N} )}{\partial \alpha_m\partial
\alpha_k}|_{\{\alpha_{i}=\alpha_{i}^{c}\}}]_{N\times N},\ee
where ${\cal V}^{d\rightarrow 0}$ is given by (\ref{pop}).\\

 It is well-known that  under the approximation (\ref{hrt}) the problem  can be reduced to  $N$ uncoupled oscillators \cite{ant} \be
{\cal H}^{HA}=\sum_{i=1}^{N}{-{1\over 2
}{d^2\over dq_{i}^2}}+{v_{i}^2 q_{i}^2 \over 2},\label{pl}\ee
where the values of $v_{i}^2$ are
 the  eigenvalues of  $\textbf{H}$. The coordinates
 $Q=(q_{1},q_{2},...,q_{N})^{T}$ (normal modes)
 are given by $Q=\textbf{U} \vec{\textbf{Z}}$, where $\vec{\textbf{Z}}=(z_{1},z_{2},...,z_{N})^{T}$, $z_{i}=x_{i}-x^{c}_{i}$,
 and $\textbf{U}$ is a matrix of  eignevectors of $\textbf{H}$.

In terms of the HA,   the asymptotic
 symmetric $(+)$ and antisymmetric $(-)$ wavefunctions of a given energy level can be constructed as follows:

\begin{eqnarray}\Psi^{(\pm)}_{g\rightarrow\infty}(x_{1},x_{2},...,x_{N})=\nonumber\\{1\over \sqrt{N!}}
\sum_{p}A_{p}\psi_{\vec{n}}(x_{p(1)}-x^{c}_{1},x_{p(2)}-x_{2}^{c},...,x_{p(N)}-x_{N}^{c})
\label{pps},\end{eqnarray}
where \begin{eqnarray} \psi_{\vec{n}}
( z_{1},z_{2},...,z_{N})=\nonumber\\=
\prod_{i=1}^N({{v_{i}}\over \pi})^{1\over 4}{1\over \sqrt{2^{n_{i}}n_{i}!}}\mathrm{e}^{-{v_{i}
q^{2}_{i}(z_{1},z_{2},...,z_{N})\over 2}}\times\nonumber\\H_{n_{i}}(\sqrt{v_{i}}q_{i}(z_{1},z_{2},...,z_{N}))\label{pp},\end{eqnarray}
 is the corresponding eigenfunction of the Hamiltonian (\ref{pl}) and $H_{n}$ is  the  $n$th  order  Hermite  polynomial.
The sum in (\ref{pps}) is over all permutations, and  $A_{p}$ is $1$ for $(+)$. For $(-)$,  $A_{p}$ is $1$ for even permutations and $-1$ for odd permutations.
In practise, the HA  works well in the regime where the localised wave packets of the particles do not overlap much, i.e., when the Wigner molecule is formed.    The general trend is that, the larger is
the value of $d$, the larger is the  value of $g$ at
which the Wigner crystal behaviour is reached.

The integral representation of the asymptotic one-particle reduced density matrix (1-RDM)  has  the form \cite{12,13}
\be
{ \Gamma}^{g\rightarrow\infty}(x,x')=\sum_{i=1}^{N}{\rho}_{i}(x,x'),\label{ex}\ee
 where the partial components ${\rho_{i}}$ are given by the following integrals
\begin{eqnarray}
\rho_{i}(x,x')=\nonumber\\{1\over N}\int_{\Re^{N-1}}(\prod_{ k\neq i}
dz_{k})\psi_{\vec{n}}(z_{1},...,z_{i-1},x-x^{c}_{i},z_{i+1},...,z_{N})\times\nonumber\\\psi_{\vec{n}}(z_{1},...,z_{i-1},x'-x^{c}_{i},z_{i+1},...,z_{N}).\label{partial}\end{eqnarray}
The Wigner crystal ground-state wavefunction that we are interested in only corresponds to the lowest energy wavefunction of  (\ref{pl});  that is,
\be \psi
 (z_{1},z_{2},...,z_{N})=\prod_{i=1}^N
({{v_{i}}\over \pi})^{1\over 4}\mathrm{e}^{-{v_{i}
q^{2}_{i}(z_{1},z_{2},...,z_{N})\over 2}}\label{pp},\ee
 In this case,   the integrals  (\ref{partial}) and  the  diagonal form of    (\ref{ex})  can be obtained  explicitly, namely  \cite{13},

\be \rho_{i}(x,x')=A_{i}\mathrm{e}^{-a_{i}[(x-x_{i}^c)^2+(x'-x_{i}^c)^2]+b_{i}(x-x_{i}^c)(x'-x_{i}^c)},\label{opo}\ee
$a_{i}, b_{i}>0$ and
\begin{eqnarray} {\Gamma}^{g\rightarrow\infty}(x,x')=\sum_{i=1}^N\sum_{l=0}^{\infty}\lambda_{l}^{(i)}
u_{l}^{(i)}(x)u_{l}^{(i)}(x'),\label{rdml1}\end{eqnarray}
respectively, where

\be
u_{l}^{(i)}({x})={\eta_{i}^{1\over 4}\over \pi^{1\over
4}\sqrt{2^l l!}}\mathrm{e}^{-{1\over 2}\eta_{i} {(x-x^{c}_{i})}^2}{H}_{l}(\sqrt{\eta_{i}}
{(x-x^{c}_{i})}),\label{fg}\ee\ and \be
\lambda_{l}^{(i)}=A_{i}\sqrt{{\pi(1-y_{i}^2)\over
\eta_{i}}}y_{i}^{l},\label{ocup}\ee
with  \be \eta_{i}=\sqrt{4a_{i}^2-b_{i}^2},\ee
\be y_{i}={{\sqrt{2a_{i}+b_{i}}-\sqrt{2a_{i}-b_{i}}}\over {\sqrt{2a_{i}+b_{i}}+\sqrt{2a_{i}-b_{i}}}}\label{44}.\ee
Note that  $\langle u_{l}^{(i)}|u_{k}^{(i)}\rangle=\delta_{lk}$ and the integral overlap  $\langle u_{l}^{(i)}|u_{k}^{(j)}\rangle$  vanishes  for any  $i\neq j$ in the limit as $g\rightarrow\infty$, where $|x_{i}^{c}-x_{j}^{c}|\rightarrow \infty$.
 The orbitals $u_{l}^{(i)}$ and the  coefficients $\lambda_{l}^{(i)}$ are  nothing  but the
 eigenvectors (natural orbitals) and  eigenvalues (occupancies) of $\Gamma^{g\rightarrow\infty}$, respectively. The values of $a_{i}$,  $b_{i}$ and $A_{i}$  satisfy $a_{i}=a_{N-i+1}$, $b_{i}=b_{N-i+1}$
  and $A_{i}=A_{N-i+1}$, which follows from the fact that  the Hamiltonian (\ref{hamk})
  is invariant  under the transformation $\vec{r}\rightarrow -\vec{r}$.
  As a result,  if $N$  is even, all the asymptotic occupancies are two-fold degenerate:
  $\lambda_{l}^{(i)}=\lambda_{l}^{(N-i+1)}$, whereas,
if $N$ is odd, the asymptotic spectrum consists of  occupancies that are two-fold degenerate and those that are
not degenerate: $\lambda_{l}^{({{N+1}\over 2})}$. 

As a matter of fact, the above    approach
 is semi-analytical as in most cases
    the equilibrium positions of the particles have to be   determined numerically. Nonetheless, the computational cost of such calculations
 is decidedly lower than that of full calculations. This provides us with  a unique opportunity
to gain insight into the properties of a Wigner molecule  formed by a large  $N$.

\section{Results}\label{results}

Now we come to the point where we  address the question of how   the properties of the  Wigner molecule ground-state  depend on $N$ and $d$. It was concluded in \cite{13} that, if $N$ and $d$ are relatively small, then,  in the expansion   (\ref{rdml1}), only the terms with $l = 0$ are substantial; that is,
 \be {\Gamma}^{g\rightarrow\infty}(x,x')\approx\sum_{i=1}^N\lambda_{0}^{(i)}
u_{0}^{(i)}(x)u_{0}^{(i)}(x'),\label{rdml28}\ee where $\lambda_{0}^{(i)}\approx N^{-1}$, which   means that  only the $N$ one-particle states (natural orbitals) are significantly occupied.   For example, for the system with  $N=3, d=1$, we found:
$\lambda_{0}^{(1)}=\lambda_{0}^{(3)}\approx0.3249$, $\lambda_{0}^{(2)}\approx0.3193$.
 However, the above issue has not yet been fully explored. Here we address it and reveal for which values of the control parameters of the system the formula (\ref{rdml28}) becomes valid.
For the present purpose,   it  is convenient to use   the so-called degree of correlation
\be \mbox{K}={P}^{-1},\ee
$P=\sum_{l}\lambda_{l}^2$,  which  counts approximately  the number of natural  orbitals actively
involved in the  expansion of the 1-RDM. This quantity is one of the transparent measures of correlation \cite{14}. In the particular case that $\lambda_{i}=1/N$, $(i=1,...,N)$, it exactly gives the number of occupied one-particle states,  $\mbox{K}=N$.

 In the perfect Wigner molecule regime   as $g\rightarrow \infty$, in which $\lambda_{l}^{(i)}=\lambda_{l}^{(N-i+1)}$, the  purity takes the form \cite{12}
\be
\mbox{P}^{g\rightarrow\infty}=2\sum_{i=1}^{N\over
 2}P_{i},\label{ee}\ee
 and \be \mbox{P}^{g\rightarrow\infty}=2\sum_{i=1}^{{N-1\over
 2}}P_{i}+P_{{N+1\over
 2}},\label{ee1}\ee
 for  even and odd values of  $N$, respectively, $P_{i}=\sum_{l}(\lambda^{(i)}_{l})^2$.
As long as $\mbox{K}^{g\rightarrow \infty}=(\mbox{P}^{g\rightarrow\infty})^{-1}\approx N$, the corresponding 1-RDM is expected to have the property  (\ref{rdml28}).
\begin{figure}[h]
\begin{center}
\includegraphics[width=0.46\textwidth]{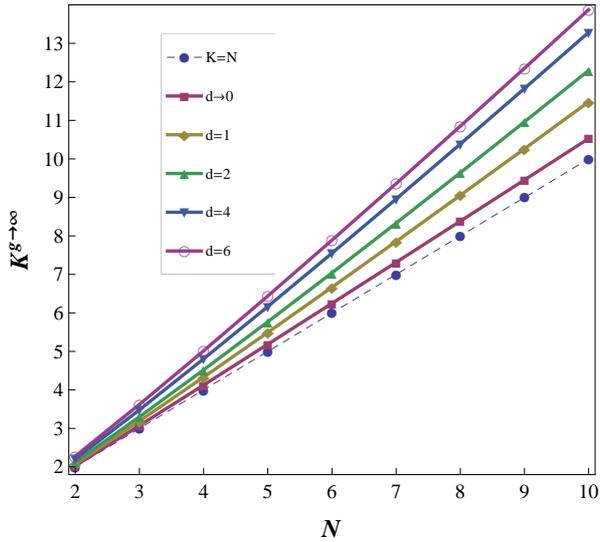}

\end{center}
\caption{\label{fffklolofghk39lg1:beh} The dependence of   $\mbox{K}^{g\rightarrow \infty}$ on $N$ for several values of  $d$.  }
\end{figure}
\begin{figure*}
\begin{center}
\includegraphics[width=0.34\textwidth]{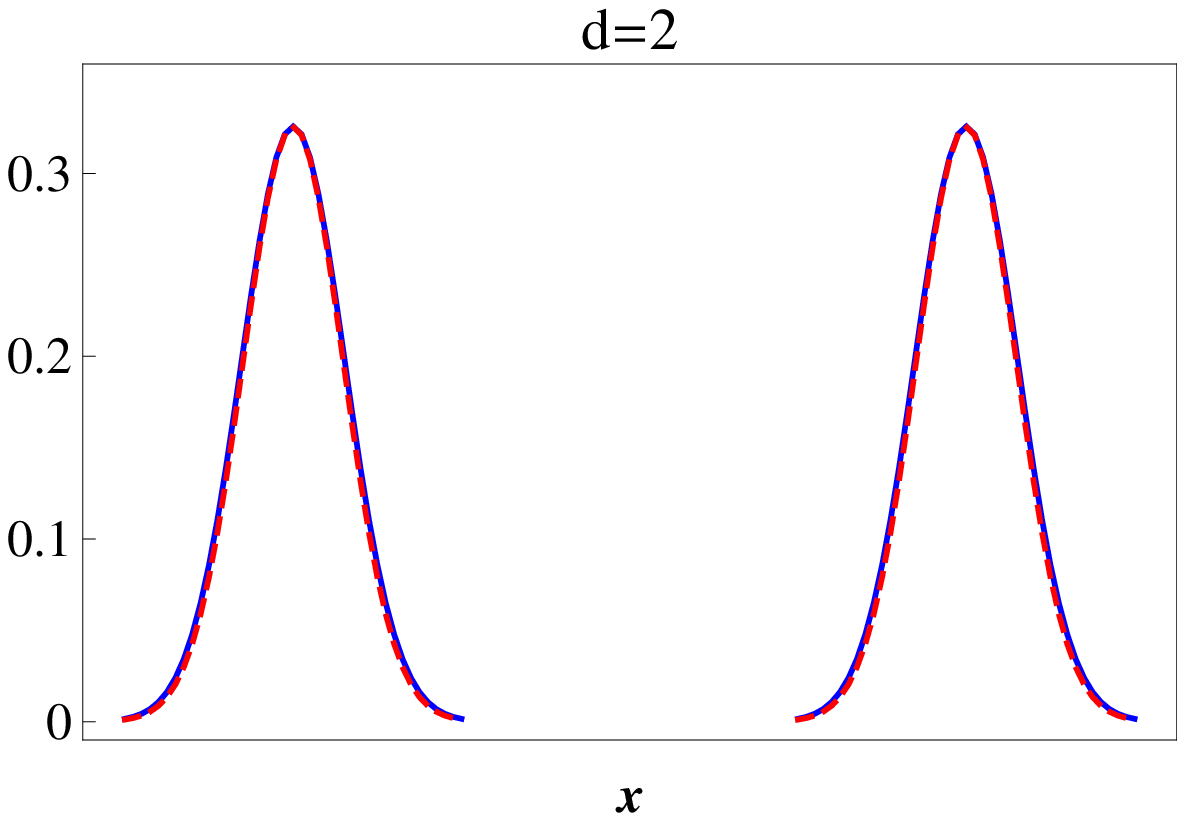}
\includegraphics[width=0.34\textwidth]{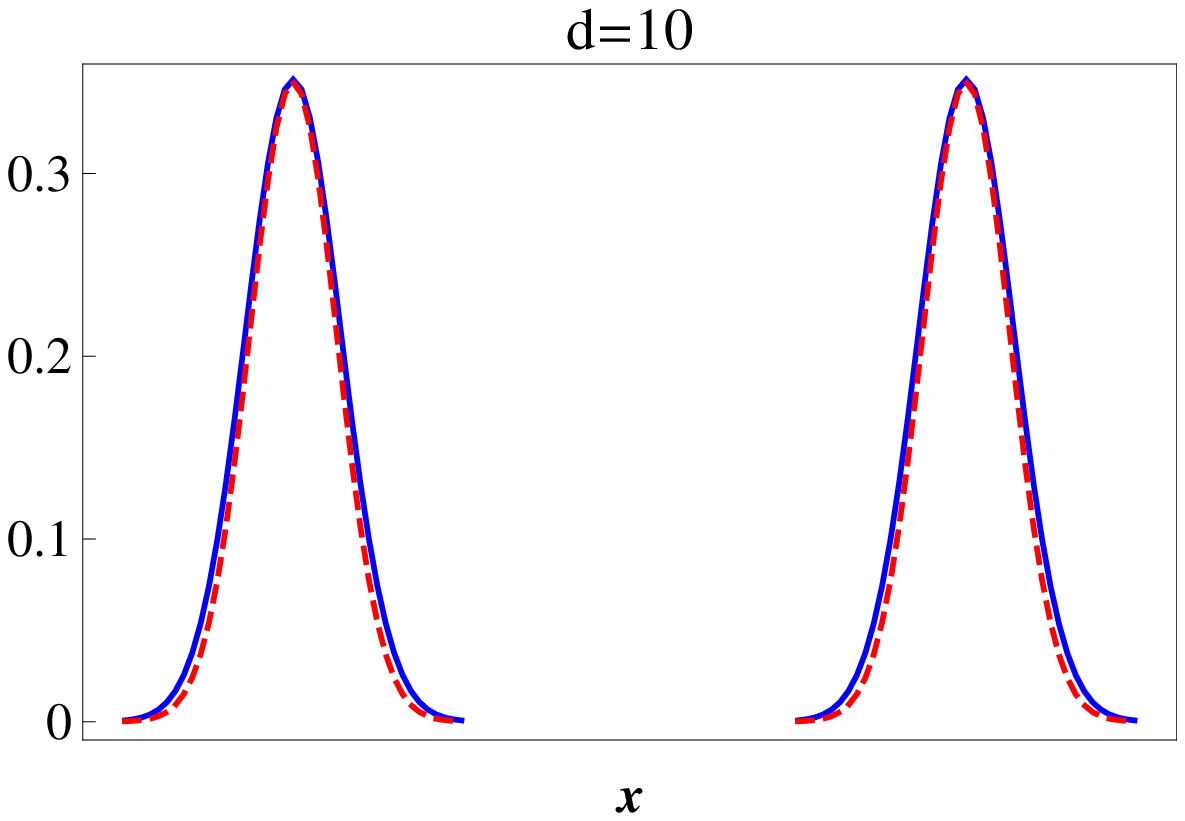}
\includegraphics[width=0.34\textwidth]{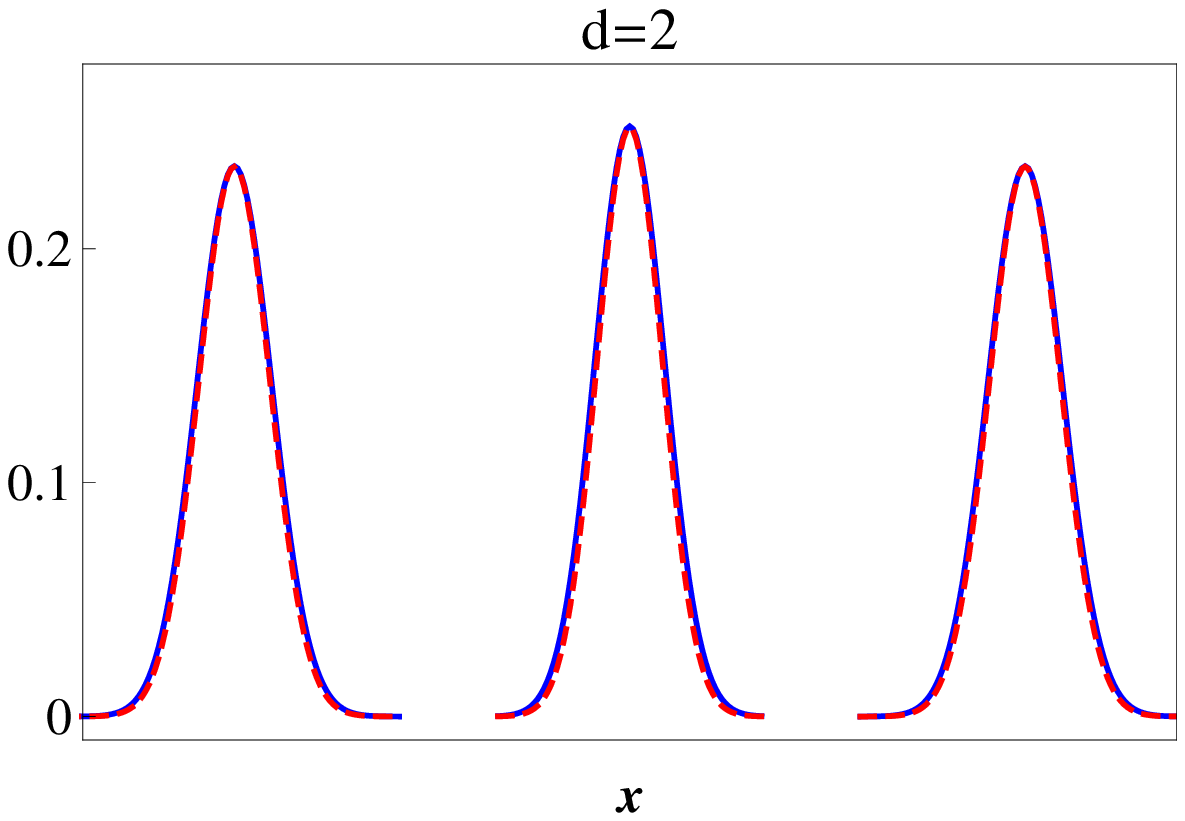}
\includegraphics[width=0.34\textwidth]{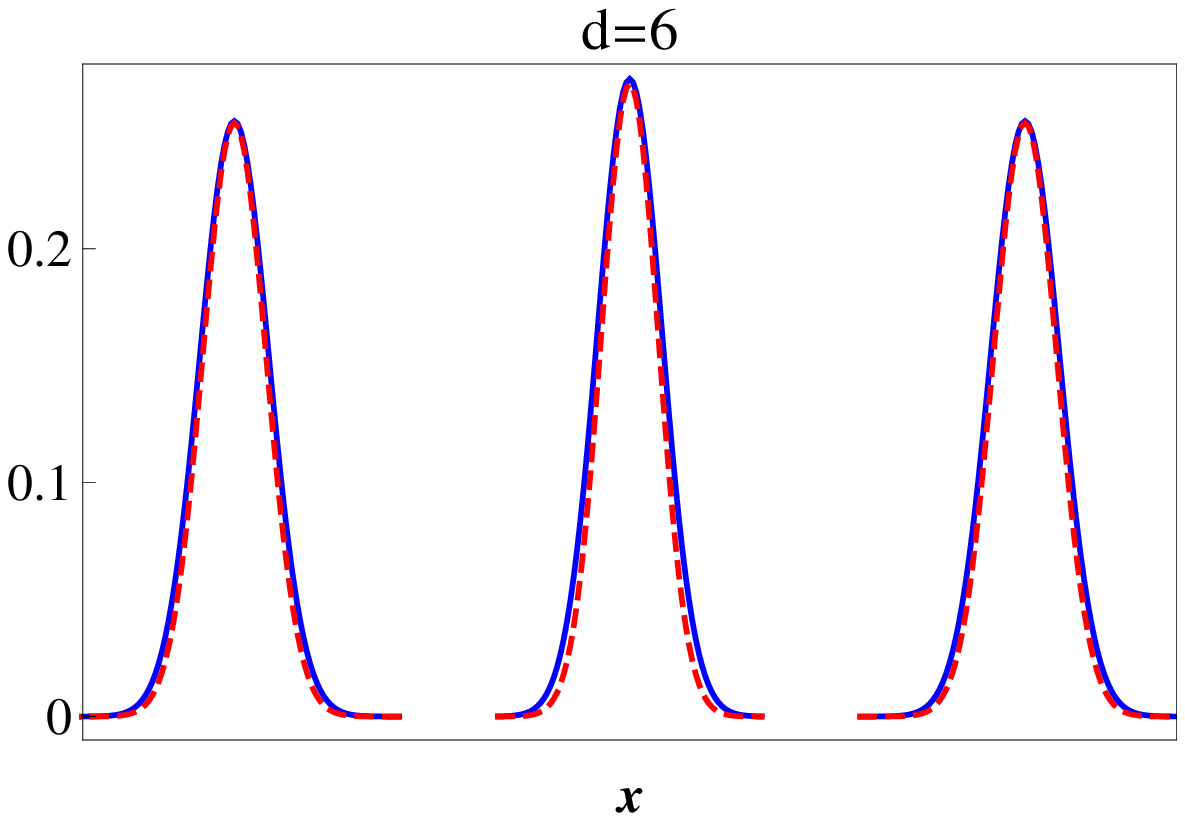}
\includegraphics[width=0.34\textwidth]{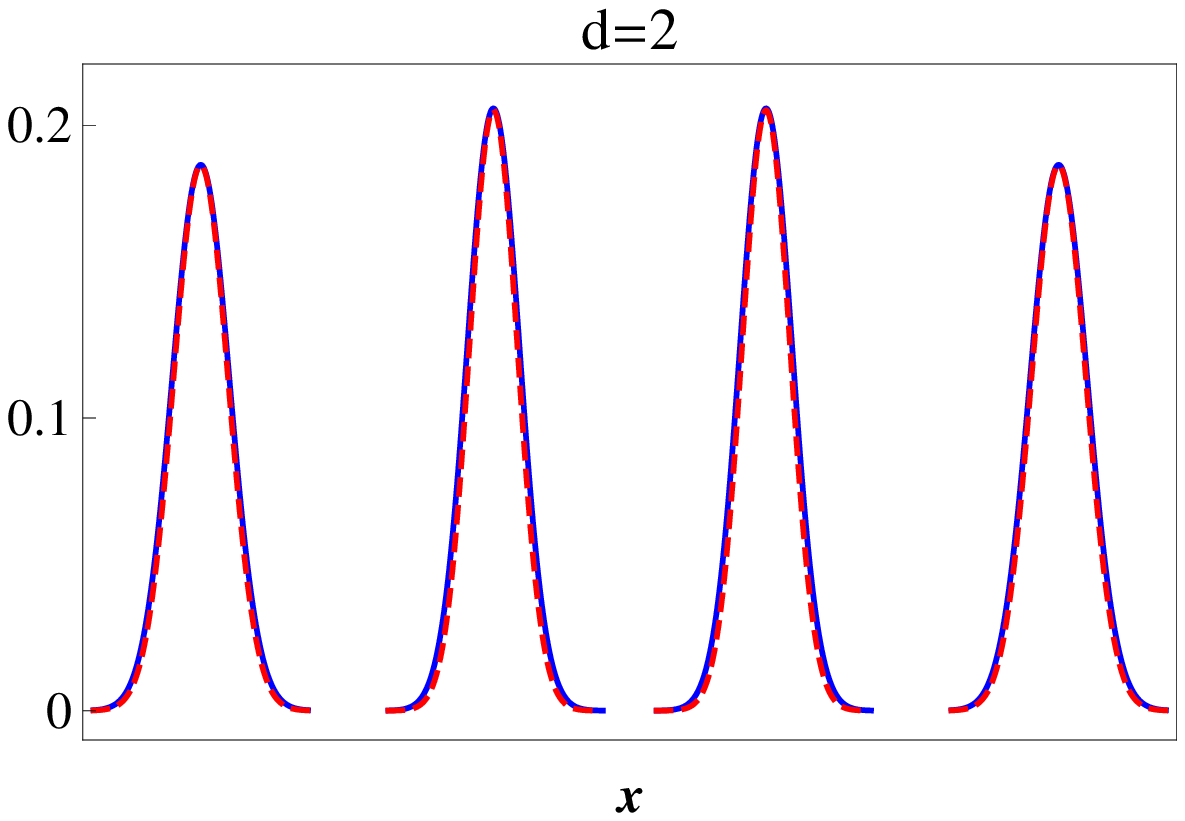}
\includegraphics[width=0.34\textwidth]{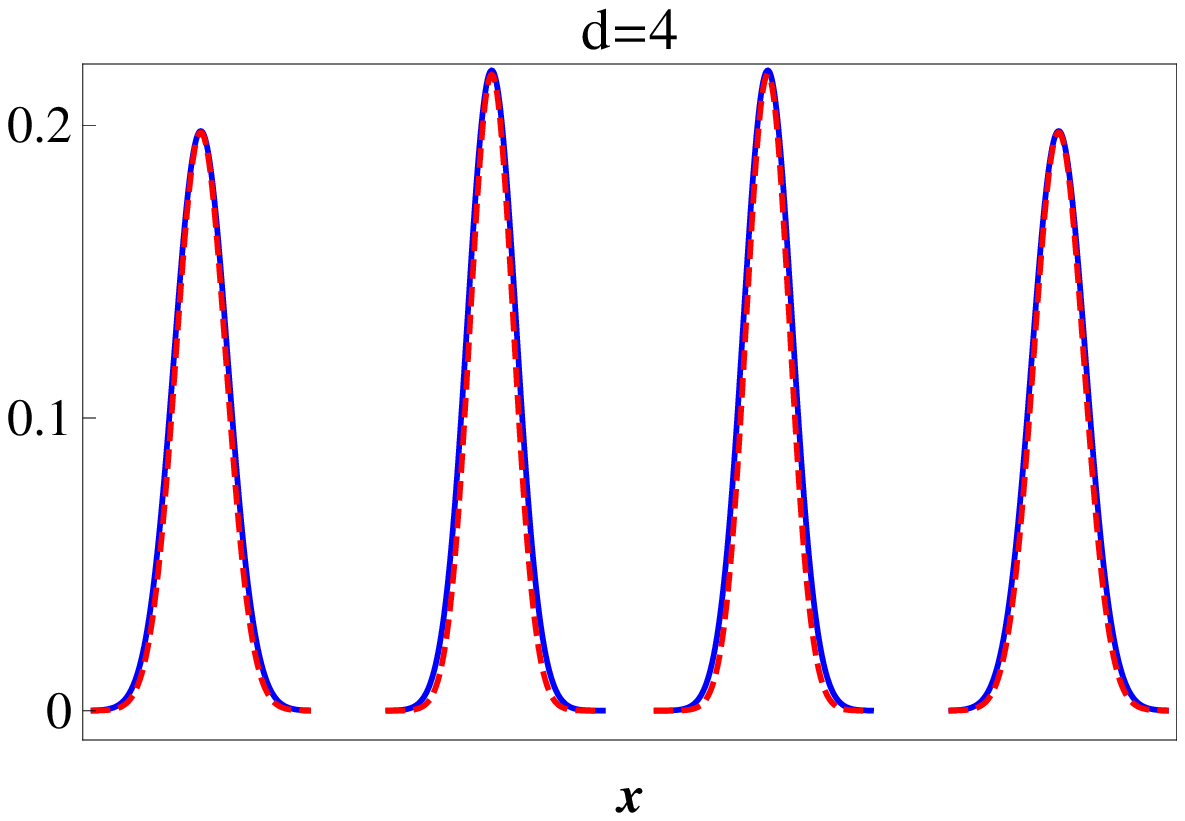}
\end{center}
\caption{\label{fffklolofghk3lg:beh}Comparison of the one-particle density of a perfect Wigner molecule ($g\rightarrow \infty$) ground-state with its approximation (\ref{rdml28}) for $N=2$:$d=2$
($\delta K=0.06$), $d=10$ ($\delta K\approx0.2$);  $N=3$:$d=2$ ($\delta K\approx0.1$), $d=6$ ($\delta K\approx0.2$) and  $N=4$:$d=2$ ($\delta K\approx0.13$), $d=4$ ($\delta K\approx 0.2$). For the sake of presentation, the density peaks are located at some fictitious points (In the limit as $g\rightarrow \infty$ the distance between the classical equilibrium positions of any pair of particles tends to infinity).}
\end{figure*}\\

Figure \ref{fffklolofghk39lg1:beh} shows the dependence of   $\mbox{K}^{g\rightarrow \infty}$ on $N$ for different
values  of $d$, including the limit of $d\rightarrow 0$.
In addition, to indicate deviations   from (\ref{rdml28}),
the values of  $\mbox{K}=N$  are  shown in this figure.
It is apparent from the results that the
value of  $\mbox{K}^{g\rightarrow \infty}$ deviates more and more from $N$ with the increase in  $N$ and/or in $d$.
This reflects the fact that, at the same time,   the natural orbitals $u_{l}^{(i)}$ with $l$ higher  than $l=0$,
that is, these beyond the approximation (\ref{rdml28}), become increasingly important.
We have estimated  that the formula (\ref{rdml28}) yields a reasonable approximation (at least to the one-particle density)  if the corresponding \textsl{relative error}   $\delta
K=(\mbox{K}^{g\rightarrow \infty}-N)/N$ does not exceed a value of about $0.2$.  Of course, the smaller is the value of $\delta K$, the better is the approximation (\ref{rdml28}).  The above is proven by the results of Figure ~\ref{fffklolofghk3lg:beh}, which compares
the  density peaks  of the perfect Wigner molecule (\ref{opo})  \be n_{i}(x)=\rho_{i}(x,x)=A_{i}\mathrm{e}^{c_{i}(x-x_{i}^c)^2},\label{opo21}\ee $c_{i}=-2a_{i}+b_{i}$
       with  their one natural orbital  approximations according to (\ref{rdml28}), $n_{i}(x)\approx\lambda_{0}^{(i)}[
u_{0}^{(i)}]^2$. The results are presented  for $N=2,...,4$ and some different  values of $d$.
As demonstrated here, the critical value of $d$ at which $\delta K=0.2$   decreases   with increasing $N$($N< N_{cr}$, where  $N_{cr}$ is the smallest value of $N$ at which   $\delta K_{d\rightarrow 0}>0.2$).

Note that the two-particle Wigner molecule formed in the limit as   $d\rightarrow 0$ is that for which the approximation (\ref{rdml28}) works the best. We obtained in this case $ \lambda_{0}^{(1,2)}=2^{5/4}/(1+2^{1/4})^2\approx0.496$, $\mbox{K}^{g\rightarrow \infty}\approx2.03$, which means that the  sum of
the remaining  asymptotic occupancies  is indeed  vanishingly small: $0.007$ \\

\begin{figure}[h]
\begin{center}
\includegraphics[width=0.48\textwidth]{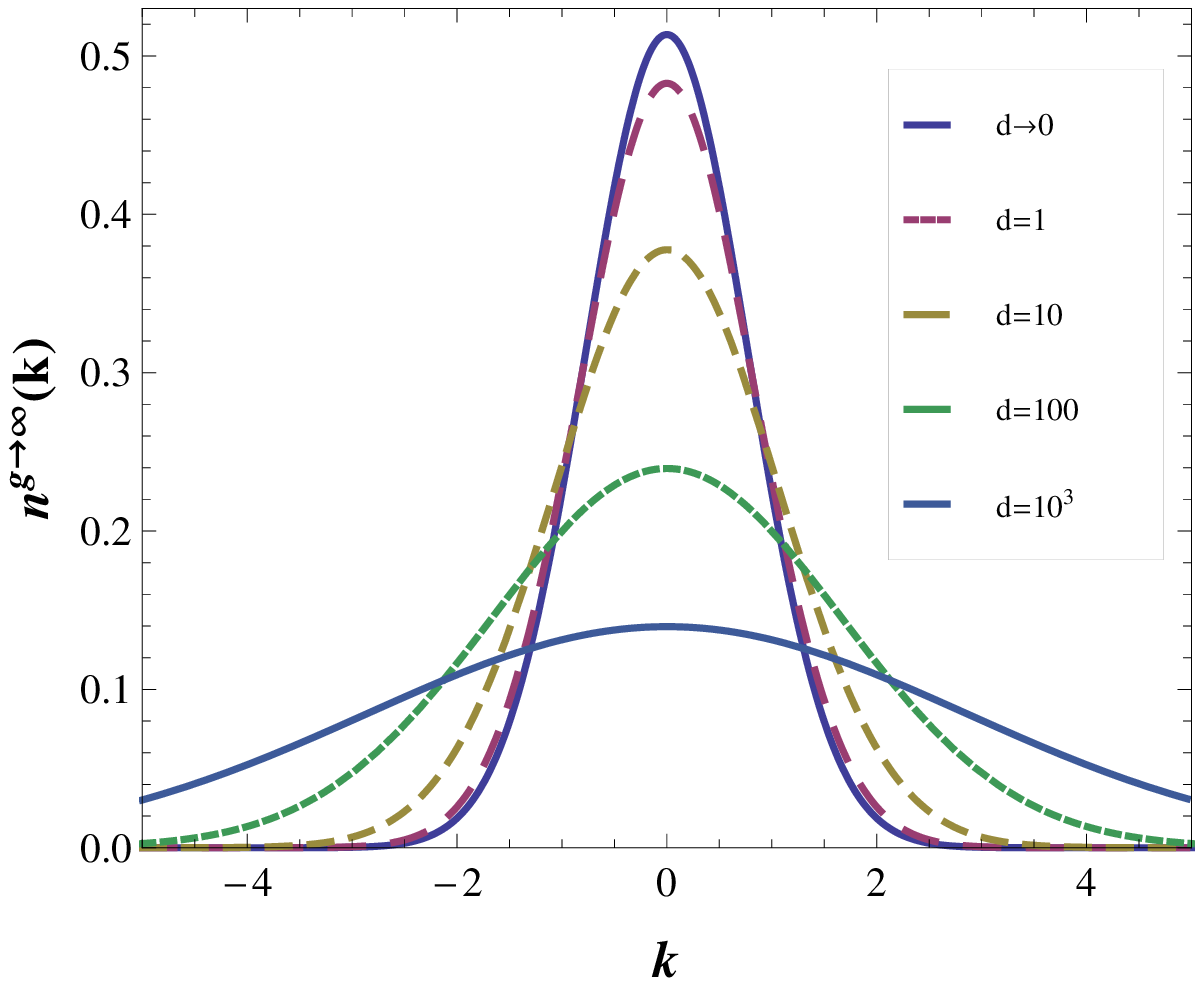}
\end{center}
\caption{\label{fffklolofgh9k39lg1:beh}  The momentum distribution for the two-particle Wigner molecule ground-state   as a function of
$d$.  }
\end{figure}
Interestingly enough, our inspection found that the partial components  (\ref{opo}) tend  to be within  the limit as $d\rightarrow \infty$ to:
\be \rho_{i}(x,x')={1\over \sqrt{\pi N}}\mathrm{e}^{-a_{i}(x-x')^2+c_{i}(x-x_{i}^c)(x'-x_{i}^c)},\label{opo1}\ee
with $a_{i}\rightarrow \infty$ and $c_{i}=-2a_{i}+b_{i}=-N$.  What follows from the above is that the corresponding 1-RDM $\Gamma^{g\rightarrow  \infty}(x,x')$  does not exhibit contributions in the off-diagonal region (note that, if $x\neq x'$, then $\mathrm{e}^{-a_{i}(x-x')^2}\rightarrow 0$ as $a_{i}\rightarrow \infty$). Accordingly, the corresponding   one-particle density peaks ($x=x'$) have the form \be n_{i}(x)={1\over \sqrt{\pi N}}\mathrm{e}^{-N(x-x_{i}^c)^2}.\label{oppo1}\ee  Further, bearing in mind Eq.  (\ref{opo1}), we immediately  conclude   that the values of $y_{i}$ in  (\ref{44}) tend to $1$ as $d\rightarrow \infty$, which means that the   clustering of the asymptotic occupancies (\ref{ocup}) around zero occurs. In other words, in this limit, the infinite degeneracy occurs, namely
   the  natural orbitals (\ref{fg}) correspond asymptotically to the same occupancy.
     Specifically, in the  case of $N=2$, which can easily be tackled
    analytically, we found that all the occupancies $\lambda^{(1,2)}_{l}$ are asymptotically equivalent
     to  $2(1/d)^{1/4}$ as $d\rightarrow \infty$, $\lambda^{(1,2)}_{l} \sim 2(1/d)^{1/4}$.

   Despite the fact that  the HA is valid only for  finite values of $d$,
  one can expect that the Wigner molecule formed in the limit of  sufficiently large   $d$   will exhibit
  approximately the  above  properties.

Finally, we show in  Figure  \ref{fffklolofgh9k39lg1:beh} the results of how the
momentum distribution \be n^{g\rightarrow \infty}(k)={1\over 2\pi}\int_{-\infty}^{\infty}\int_{-\infty}^{\infty}\Gamma^{g\rightarrow  \infty}(x,x') \mathrm{e}^{- i k  (x-x')}dx dx' \label{jk}\ee of
the two-particle Wigner molecule ground-state changes with  $d$.
 In this case,  the integrals (\ref{jk}) can be performed explicitly:   \be n^{g\rightarrow \infty}(k)=\frac{\sqrt{\frac{2}{\pi }} \mathrm{e}^{-\frac{2 \left(\sqrt{d+2}-1\right)
   k^2}{d+1}}}{\sqrt{\sqrt{d+2}+1}}.\ee  As can be seen,  the
momentum distribution  broadens as the value of $d $  increases, which reflects  the fact that, at the same time,  a reduction of density in the
off-diagonal regions of $\Gamma^{g\rightarrow  \infty}$ occurs. Obviously, for   larger $N$, the similar behaviour is expected.

\section{Conclusions}\label{con}
Within the framework of the  HA, we studied the properties  of the Wigner molecule
 formed by $N$ identical particles confined in a harmonic trap and interacting via the inverse power-law
 potential $ |x|^{-d}$. Among other things,   we elaborated the scheme to study
     the special case  as $d$ approaches zero.  Our results show
 the dependencies of the degree of correlation  on  $d$  for the $N$- particle Wigner molecule ground-state
 in a wide range of values of $N$. As a general trend we found that
  the number of natural orbitals contributing   to the correlation of such a  state grows with increasing  $d$.
 In addition, we showed that the 1-RDM of  the  Wigner molecule  formed  in the limit as  $d$ approaches $\infty$  does not  exhibit contributions in the off-diagonal regions. The corresponding  one-particle
density has been  derived  in a closed analytic form in the general case of $N$ particles.  It turned out that it exhibits the  peaks of
 the same profile.

\bibliography{aipsamp}

\end{document}